\documentclass{article}
\usepackage{glas}
\usepackage{epsfig}
\begin{document}
\begin{titlepage}{GLAS-PPE/97--14}{December 1997}
\title{LHC-B Ring Imaging Cherenkov Detector}
\author{N.\ H.\ Brook}
\collaboration{on behalf of the LHC-B collaboration}
\begin{abstract}
The progress towards the realisation of the LHC-B Ring Imaging Cherenkov
detector is reported.
\end{abstract}
\vfill
\conference{invited talk given at the \\
 5th international workshop on
B-Physics at hadron machines (Beauty'97), \\
Los Angeles, California, USA. \\Oct 13-17, 1997}
\end{titlepage}
\section{Introduction}
This paper reports on the development of the Ring Imaging 
Cherenkov (RICH) detector for LHC-B.
The LHC-B experiment is a proposed single arm spectrometer for the LHC
optimised for $B$ physics.
Particle identification is crucial for the study of $B$-physics and
CP violation at the LHC. 
The many $B$-meson decay modes to be
studied makes it  necessary to have $\pi/K$ separation over a momentum
range of $1 < p < 150 {\rm\ GeV/c}.$
This reduces the background of the selected final state and,
in addition, provides an efficient flavour tag of the signal $B$-meson.
The overall concept and performance of LHC-B is discussed in more detail
elsewhere in these proceedings~\cite{nev}.

\begin{figure}[hbt]
\centerline{
\epsfig{figure=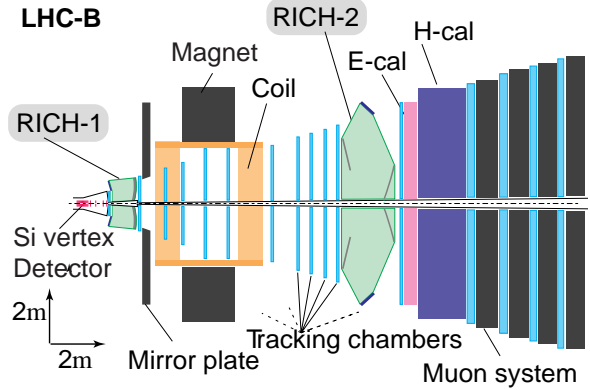,bbllx=40pt,bblly=220pt,bburx=545pt,bbury=565pt,
width=8.0cm}}
\caption{The LHC-B spectrometer}
\label{fig:expt}
\end{figure}

\begin{figure*}[htb]
\centerline{
\epsfig{figure=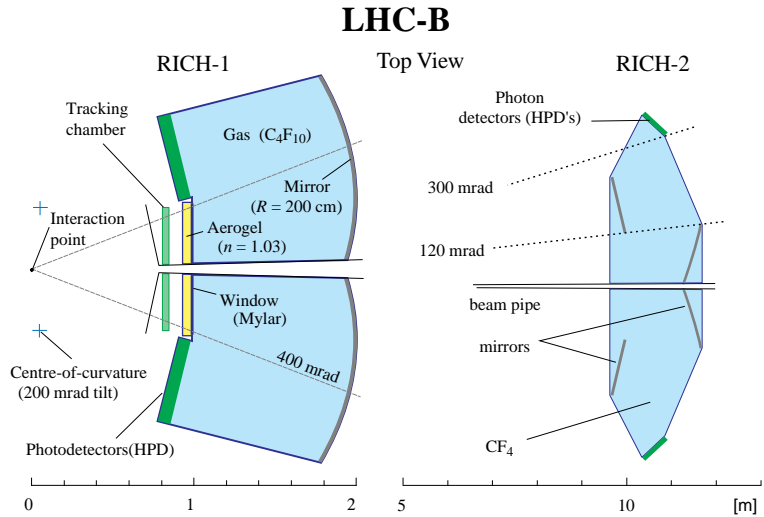,width=10.0cm}}
\caption{Layout of the LHC-B RICH detectors(note different scales).}
\label{fig:RICH}
\end{figure*}

RICH detectors meet the experimental criteria
for particle identification in
 the required momentum range.
It is proposed to use two
RICH detectors in the LHC-B experiment, see fig~\ref{fig:expt}.
The upstream detector (RICH1), fig.~\ref{fig:RICH},
has a combined gas and aerogel radiator and is situated in
front of the dipole magnet. The aerogel radiator is placed against the
entrance window of the second gaseous ($\rm C_4F_{10}$) radiator.
A spherical mirror with a radius of curvature of 190~cm is tilted by $\approx
250$ mrad to reflect the Cherenkov light onto an array of photodetectors
situated outside the experimental acceptance. The downstream RICH 
(RICH2), fig.~\ref{fig:RICH}, uses $\rm CF_4$ as its radiator with a spherical
mirror with a radius of curvature of 820~cm tilted by 370 mrad. An
additional flat mirror is tilted by 240 mrad  to bring the
Cherenkov photons out of the acceptance of the experiment.

Each RICH detector has two photodetector planes giving a total area
of $2.9\ \rm m^2.$ The chosen detector technology must have a high
quantum efficiency, a spatial resolution of at least $2.5 \times 2.5 \
{\rm mm^2}$ and to have a fast readout consistent
with 25~ns bunch crossing of the LHC.
Two candidate technologies exists~\cite{rusack}: 
hybrid photodiodes (HPDs)
and multianode photomultipliers. These detectors are commercially
available but not in designs that meet the experimental needs of 
LHC-B.
A program of R\&D is
currently underway on the development of a HPD with a large active area.

\section{HPD development}
Two complimentary approaches to the development of the HPDs
are being investigated: the `pixel HPD', which uses the
bump bonding of a silicon pixel detector to a readout chip,
and the `pad HPD' where a silicon pad detector is readout using routing
lines wire-bonded to front-end chips around its circumference.

\subsection{Pixel HPD}
The pixel HPD is being developed in close
collaboration with DEP~\cite{DEP}. It is based on standard image
intensifier technology that strongly focuses
 photoelectrons onto a segmented silicon
pixel array. The feasibility of this approach has already been
demonstrated with the `IPSA-tube'~\cite{IPSA}.
Work is ongoing on a scale model prototype with an active photocathode
diameter of 40~mm and an anode diameter of 11~mm. The anode is assembled
with the LHC1 chip developed at CERN~\cite{LHC1}. This chip contains a
detector array of 2048 pixels, $50 \times 500 \mu{\rm m^2}$ in size and
it has a
lowest achievable comparator threshold of $4000e^-$ with a spread of $1000e^-.$
Unfortunately these detector properties are not compatible with the
needs of LHC-B.
In particular, the pixel dimension $(50 \mu{\rm m})$ 
is too small compared to the RICH granularity ( $500 \times 500 \mu{\rm m^2}).$
Nor are the the threshold properties of the LHC1
 chip compatible with
LHC-B requirements, though
recent development in pixel electronics have achieved
comparator thresholds down to $1400e^-$ with an RMS of $90e^-.$
The specific needs of LHC-B are being investigated.

\subsection{Pad HPD}
The pad HPD will be housed in a cylindrical glass envelope capped with
a UV-glass entrance window with a $\rm K_2CsSb$ photocathode. 
A visible light photocathode deposition facility that allows a
high vacuum
seal of the HPD baseplate onto the metal flange of the glass envelope
has been designed. The final
assembly of this apparatus was performed at CERN at the end 
of 1997. The signal from the photoelectrons in the silicon sensor will be
detected by front end chips 
placed around the edge of the sensor via wires
bonded to the routing lines from the pad. Focussing electrodes, fixed
in the glass envelope, will demagnify the image by a factor of $2.3.$ The
2048 pad silicon detector contains pads of dimensions $ 1\time 1 {\rm
mm^2}$. Successful tests have been performed on the sensor with
photoelectrons up to 20 keV with (the non LHC speed) VA3 chip~\cite{VA3}.
A signal/noise ratio of $\approx 10$ has been achieved. Earlier tests
with a 256 pad sensor had achieved a signal/noise ratio of $\approx 18.$
This difference is under investigation, but one possible reason for this
degradation in performance is differences in the manufacturing of the
wafer.

The SCT-128A~\cite{SCT128} analogue chip which was developed for the 
ATLAS silicon tracker is being modified to achieve a noise level of
$\sim 600 e^-.$ Additional modifications will be needed to meet the
requirements of the pad HPD for LHC-B, in particular the multiplexing
properties of the chip.

\begin{figure}[hbt]
\centerline{
\epsfig{figure=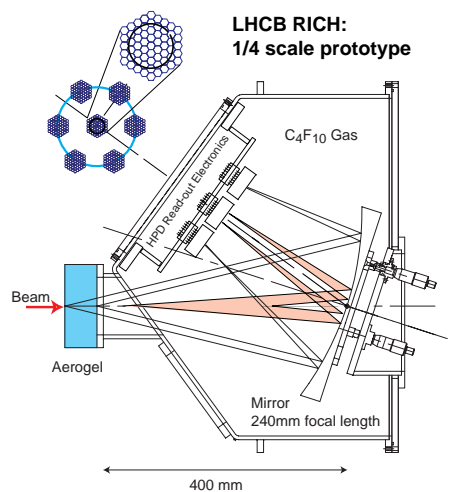,bbllx=10pt,bblly=100pt,bburx=560pt,bbury=700pt,
width=6.0cm}}
\caption{Configuration 1: the $1/4$ scale prototype RICH vessel}
\label{fig:layout}
\end{figure}

\section{RICH Prototype}

A prototype of the downstream RICH detector
was tested in the T9 test beam at the CERN SPS
during the Spring and Summer of 1997. A planar array of
seven 61-pixel HPD's from DEP
were used to detect the Cherenkov photons produced in
aerogel, air and $\rm C_4F_{10}$ radiators. In configuration-1,
fig.~\ref{fig:layout}, of the prototype
the light is focussed by a 240 mm focal length mirror which corresponds
to a $1/4$ scale of the RICH1 detector. A full scale prototype
(configuration 2) was also used which has a 1143~mm mirror to focus
rings from $\rm C_4F_{10}$ onto an array of six 61-pixel HPD's.
This was achieved by adding extension arms to configuration 1.

The 61 pixel HPD has a silicon diode detector
segmented as a hexagonal array with pad dimensions of 2~mm face-to-face.
The HPD was operated at a high voltage of 12kV. Using a pulsed light
emitting diode the complete readout and data acquisition chain was tested.
The pedestal, the single, double and triple photoelectrons peaks were
clearly visible with a signal/noise ratio of $\approx 5.7.$ Most of
this noise is associated with the input capacitance of the feedthrough
and printed circuit boards.


The test beam provides charged particles of either polarity and the
momentum can be tuned in the range $2-15.5 \rm\ GeV/c.$ The particle
type is identified by measuring the signal pulse height from a $\rm
CO_2$ threshold Cherenkov counter installed 30~m upstream from the
prototype. The prototype vessel was aligned with the beam axis. Charged
particles which provide the trigger are selected using using
scintillation counters, two upstream and two downstream of the vessel.
A photoelectron hit is defined to be a HPD pixel with a signal pulse
height $4\sigma$ above the pedestal mean, where $\sigma$ is the rms width
of the pedestal peak.

\begin{figure}[htb]
\centerline{
\epsfig{figure=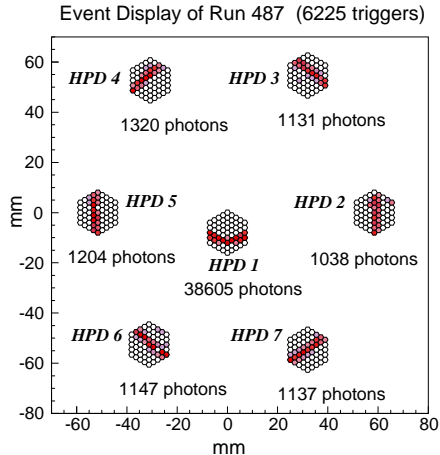,bbllx=10pt,bblly=100pt,bburx=555pt,bbury=670pt,
width=6.0cm}}
\caption{An event display from aerogel and $\rm C_4F_{10}$ radiators in
RICH configuration 1, integrated over run 487.}
\label{fig:rings}
\end{figure}

Using RICH configuration 1, data were taken with a 10 GeV/c negatively
charged beam with 18~mm thickness of aerogel.
Fig.~\ref{fig:rings} shows an arc of a ring on the central HPD, whose
radius is compatible with that expected from $\rm C_4F_{10}.$ The outer
HPD's clearly exhibit a ring which originates from the aerogel radiator.

The number of photoelectrons per triggered event was measured for all
three radiators in the vessel. 
For this analysis a threshold of $3\sigma$ was set for
individual pixels and multiple photoelectrons were taken into account.
The mean number of photoelectrons are shown in table~\ref{tab:photons}.
The partial geometrical coverage of the aerogel and gas rings was
calculated from simulation with $\approx 5\%$ uncertainity.
The expected photoelectron yields was calculated from simulation
which included the properties of the aerogel, mirror and photocathode
efficiencies. The overall precision in the expected yield is estimated
to be 15\%.
The comparison between observed and expected  yields are given in
table~\ref{tab:photons}. The numbers from this preliminary analysis are
compatible within 30\%.

\begin{table}[htb]

\centering
\begin{tabular}{|l||c|c|c||c|}
\hline
 Radiator &  Raw &  Bkg. &  Eff. &  Ratio \\
 &  hits &  corr. &  corr. & \\ \hline
 Air & 4.92 & 4.56 & 4.80 & 0.99 \\
$\rm C_4H_{10}$ & 7.85 & 7.49 & 33.55 & 1.07 \\
 Aerogel & 1.79 & 1.31 & 10.71 & 0.72 \\
\hline
\end{tabular}

\caption{Observed number of photoelectrons per event for air, aerogel
and $C_4F_{10}$ radiators. The columns give the number of raw hits, the
numbers after correction for background and for geometrical efficiency,
and the comparison with the expected yield.}
\label{tab:photons}
\end{table}

\begin{figure}[hbt]
\centerline{
\epsfig{figure=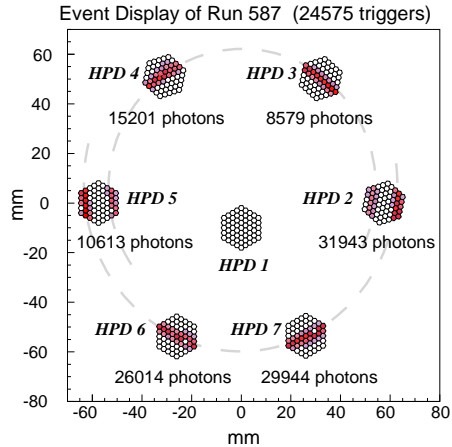,bbllx=45pt,bblly=120pt,bburx=580pt,bbury=655pt,
width=6.0cm}}
\caption{An event display showing $\pi/K$ separation, using $\rm C_4F_{10}$ 
radiator in RICH configuration 2, integrated over run 587.}
\label{fig:config2}
\end{figure}

The full scale RICH1 prototype was studied using configuration-2. The
longer focal length of the mirror means the $\rm C_4F_{10}$ ring now
spans the outer 6 HPDs. The event display shown in
fig.~\ref{fig:config2} is obtained from negatively charged 15.5 GeV/c
momentum beam. The $K:\pi$ ratio of the triggering particles has been
enhanced to $1:2$ using the threshold Cherenkov counter.
Fig.~\ref{fig:config2} shows segments of two rings; an inner ring from
the incident kaons and an outer ring from the pions. It can be seen that
the number of hits observed in HPD 3 is lower than in HPD 4. (Similarly
HPD 5 has fewer hits than HPD 2.) This is because HPD's 3 and 5 have
mylar windows in front of their photocathodes which absorb the UV photons.

\section{Summary}
The major outstanding issue for the LHC-B RICH detector is the
demonstration of a photodetector that matches LHC-B requirements.
For the pixel HPD it is envisaged to produce and test a 80~mm diameter tube
whilst a pixel chip is developed in parallel that meets the experimental
needs. Test in 1998 are planned for the pad HPD using the 2048 pad
detector under vacuum 
with the designed focussing and then eventually with a bialkali photocathode.

The RICH prototype tests have been successful. Clear Cherenkov rings
from gas and aerogel radiators have been observed for the first time 
using HPD's as photon detectors. The preliminary
measured photon yields are compatible within 30\% of expectations 
based on simulations.
Further analysis of the data is investigating the reconstruction
resolution of the Cherenkov angle for each recorded hit from both
the gas and aerogel radiators.
Further prototype testbeam runs are being planned to study RICH 2. It is
also planned to use the RICH prototype to test the various
photodetectors as they become available.


\begin{thebibliography}{9}

\bibitem{nev}
N.~Harnew, \newblock LHC-B B-Physics Performance, \emph{to appear in these
proceedings.}

\bibitem{LOI}
LHC-B Collaboration, LHC-B Letter of Intent, CERN/LHCC 95-5.

\bibitem{rusack}
R.~Rusack, \newblock Advances in Photon Detectors, \emph{to appear in these
proceedings.}

\bibitem{DEP}
Delft Electronic Products BV, 9300AB Roden, Netherlands.

\bibitem{IPSA} 
T. Gys et al, \emph{Nucl. Instr. Meth.\/} \textbf{A355} (1995) 386.

\bibitem{LHC1}
E. Heijne et al, \emph{Nucl. Instr. Meth.\/} \textbf{A383} (1996) 55.

\bibitem{VA3} P. Weilhammer et al, \emph{Nucl. Instr. Meth.\/}
\textbf{A383} (1996) 89.

\bibitem{SCT128}
S. Anghinolfi et al, \emph{IEEE Trans. Nucl. Sci.\/} \textbf{44} (1997)
298.

\bibitem{NIM}
E. Albrecht et al, First Observation of Cherenkov Ring Images using
Hybrid Photon Detectors, submitted to \emph{Nucl. Instr. Meth.}

\end{thebibliography}
\end{document}